\documentclass[sigconf]{acmart}

\AtBeginDocument{%
  \providecommand\BibTeX{{%
    \normalfont B\kern-0.5em{\scshape i\kern-0.25em b}\kern-0.8em\TeX}}}

\setcopyright{rightsretained}
\copyrightyear{2021}
\acmYear{2021}
\acmDOI{}
\acmConference[MIS2'21]{MIS2'21: Misinformation and Misbehavior Mining on the Web Workshop held in conjunction with KDD 2021}{August 15, 2021}{Online}
\acmBooktitle{MIS2'21: Misinformation and Misbehavior Mining on the Web Workshop held in conjunction with KDD 2021, August 15, 2021, Online}
\acmPrice{}
\acmISBN{}

\usepackage{lipsum}
\usepackage[flushleft]{threeparttable} 

\begin{document}

\title{A preliminary approach to knowledge integrity risk assessment in Wikipedia projects}

\author{Pablo Aragón}
\affiliation{
  \institution{Wikimedia Foundation}
  \city{Barcelona}
  \country{Spain}
}
\email{paragon@wikimedia.org}

\author{Diego Sáez-Trumper}
\affiliation{
  \institution{Wikimedia Foundation}
  \city{Barcelona}
  \country{Spain}
}
\email{diego@wikimedia.org}

\renewcommand{\shortauthors}{Aragón and Sáez-Trumper.}

\begin{abstract}
Wikipedia is one of the main repositories of free knowledge available today, with a central role in the Web ecosystem. For this reason, it can also be a battleground for actors trying to impose specific points of view or even spreading disinformation online. There is a growing need to monitor its ``health'' but this is not an easy task. Wikipedia exists in over 300 language editions and each project is maintained by a different community, with their own strengths, weaknesses and limitations. In this paper, we introduce a taxonomy of knowledge integrity risks across Wikipedia projects and a first set of indicators to assess internal risks related to community and content issues, as well as external threats such as the geopolitical and media landscape. On top of this taxonomy, we offer a preliminary analysis illustrating how the lack of editors' geographical diversity might represent a knowledge integrity risk. These are the first steps of a research project to build a Wikipedia Knowledge Integrity Risk Observatory. 
\end{abstract}


\maketitle

\section{Introduction}

The Web has become the largest repository of knowledge ever known in just three decades. However, we are witnessing in recent years the proliferation of sophisticated strategies that are heavily affecting the reliability and trustworthiness of online information. Web platforms are increasingly encountering misinformation problems caused by deception techniques such as astroturfing~\cite{zhang2013online}, harmful bots~\cite{ferrara2016rise}, computational propaganda~\cite{woolley2018computational}, sockpuppetry~\cite{kumar2017army}, data voids~\cite{golebiewski2018data}, etc. 

Wikipedia, the world’s largest online encyclopedia in which millions of volunteer contributors create and maintain free knowledge, is not free from the aforementioned problems. Disinformation is one of its most relevant challenges~\cite{saez2019online} and some editors devote a substantial amount of their time in patrolling tasks in order to detect vandalism and make sure that new contributions fulfill community policies and guidelines~\footnote{\url{https://en.wikipedia.org/wiki/Wikipedia:Policies_and_guidelines}}~\cite{morgan2019research}. Furthermore, \emph{Knowledge Integrity} is one of the strategic programs of Wikimedia Research with the goal of identifying and addressing threats to content on Wikipedia, increasing the capabilities of patrollers, and providing mechanisms for assessing the reliability of sources~\cite{zia2019}.

Many lessons have been learnt from fighting misinformation in Wikipedia~\cite{kelly2021} and analyses of recent cases like the 2020 United States presidential election have suggested that the platform was better prepared than major social media outlets~\cite{morrison2020}. However, there are Wikipedia editions in more than 300 languages, with very different contexts. To provide Wikipedia communities with an actionable monitoring system, this paper introduces a preliminary approach consisting of a taxonomy of knowledge integrity risks in Wikipedia projects, based on a review of the state of the art literature, and an initial set of indicators to be the core of a Wikipedia Knowledge Integrity Risk Observatory.

\section{Taxonomy of knowledge integrity risks}

\begin{figure}[h!]
  \centering
  \includegraphics[trim={0cm 0 0cm 0}, clip, width=\linewidth]{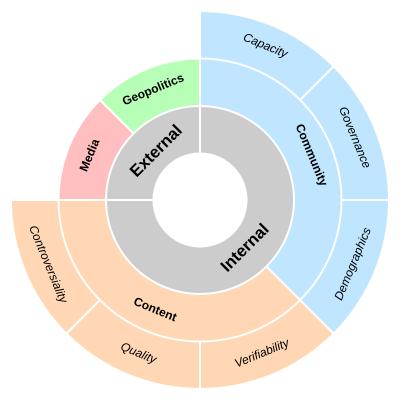}
  \caption{Taxonomy of knowledge integrity risks in Wikipedia.}
  \label{fig:taxonomy}
\end{figure}

Risks to knowledge integrity in Wikipedia can arise in many and diverse forms. Inspired by a recent work that has proposed a taxonomy of knowledge gaps for Wikimedia projects~\cite{redi2021taxonomy}, we have reviewed works by the Wikimedia Foundation, academic researchers and journalists that provided empirical evidence of knowledge integrity risks. Then, we have classified them by developing a hierarchical categorical structure that is presented in Figure~\ref{fig:taxonomy}.

We initially differentiate between \textbf{internal} and \textbf{external} risks according to their origin. The former correspond to issues specific to the Wikimedia ecosystem while the latter involve activity from other environments, both online and offline. 
For internal risks, we have identified the following categories focused on either \textbf{community} or \textbf{content}:

\begin{itemize}
  \setlength\itemsep{0.5em}

    \item \textbf{Community capacity}: Pool of resources of the community. Risks have been found when, given the size of the corresponding Wikipedia project (volume of articles, edits and editors), there is a shortage of patrolling resources, i.e., editors with elevated user rights (e.g., admins, checkusers, oversighters)~\cite{saez2019online, morgan2019research,song2020, sato2021} and tools (e.g., specialized bots, scripts, MediaWiki extensions, desktop/web apps)~\cite{saez2019online,morgan2019research}.
    
    \item \textbf{Community governance}: Situations and procedures involving decision-making within the community. The reviewed literature has identified risks like the unavailability of local rapid-response noticeboards on smaller wikis~\cite{morgan2019research} or the abuse of blocking practices by admins~\cite{rogers2012neutral, sato2021, song2020}.
    
    \item \textbf{Community demographics}: Characteristics of community members. Some analyses highlight that the lack of geographical diversity might favor nationalistic biases~\cite{rogers2012neutral, sato2021}. Other relevant dimensions are editors' age and activity since misbehavior is often observed in editing patterns of newly created accounts~\cite{kumar2015vews,kumar2016disinformation,joshi2020detecting} or accounts that have been inactive for a long period to avoid certain patrolling systems or that are no longer monitored and became hacked~\cite{morgan2019research}.

    \item \textbf{Content verifiability}: Usage and reliability of sources in articles. This category is directly inspired by one the three principal core content policies of Wikipedia (WP:V)~\footnote{\url{https://en.wikipedia.org/wiki/Wikipedia:Verifiability}} which states that readers and editors must be able to check that information comes from a reliable source. It is referred to in several studies of content integrity~\cite{redi2019citation,lewoniewski2019multilingual,saez2019online}. 

    \item \textbf{Content quality}: Criteria used for the assessment of article quality. Since each Wikipedia language community decides its own standards and grading scale, some works have explored language-agnostic signals of content quality such as the volume of edits and editors or the appearance of specific templates~\cite{lewoniewski2017relative,lewoniewski2019multilingual}. In fact, there might exist distinctive cultural quality mechanisms as these metrics do not always correlate with featured status of articles~\cite{rogers2012neutral}.
    
    \item \textbf{Content controversiality}: Disputes between community members due to disagreements about the content of articles. Edit wars are the best known phenomenon that occurs when content becomes controversial~\cite{yasseri2014most, rogers2012neutral}, sometimes requiring articles to be protected~\cite{spezzano2019detecting}. 
\end{itemize}

For external risks, we have identified the following categories:

\begin{itemize}
  \setlength\itemsep{0.5em}
    \item \textbf{Media} References and visits to the Wikipedia project from other external media on the Internet. Unusual amount of traffic to specific articles coming from social media sites or search engines may be a sign of coordinated vandalism~\cite{morgan2019research}.
    
    \item \textbf{Geopolitics}: Political context of the community and content of the Wikipedia project. Some well resourced interested parties (e.g., corporations, nations) might be interested in externally-coordinated long-term disinformation campaigns in specific projects~\cite{morgan2019research,shubber2021}.

\end{itemize}

\section{Initial set of indicators and preliminary analysis}

To capture risks in each category of the proposed taxonomy, we have compiled an initial set of indicators presented in Table~\ref{tab:indicators}. The criteria for proposing these indicators are that they should be simple to be easily interpreted by non-technical stakeholders, comparable across wikis, language-agnostic and periodically updatable. For this reason, they are counts of items (e.g., articles, editors, edits, etc.) or distributions of items over informative features. 

\begin{table*}[]
\centering
\caption{Initial set of indicators for a Wikipedia Knowledge Integrity Risk Observatory.}
\label{tab:indicators}
\begin{tabular}{l|p{40em}}
\toprule
\textbf{Risk category}     & \textbf{Candidate indicators} \\ \midrule
\midrule

Community capacity & 
Number of articles, editors, active editors, editors with elevated user rights (admins, bureaucrats, checkusers, oversighters, rollbackers); ratio of active editors with elevated user rights; number of specialized patrolling tools; number of AbuseFilter rules.\\
\midrule

Community governance & 
Number of requests in steward's noticeboard; number of global stewards knowledgeable with that language; number of requests for comment (local and meta); ratio of articles for deletion; ratio of blocked accounts (spam, long-term abuse, etc.).\\
\midrule

Community demographics   &
Distribution of views and edits by country; distribution of active editors by age, local activity and cross-wiki activity.\\
\midrule
\midrule

Content verifiability    &
Distribution of articles by number of citations, number of scientific citations and number of citation and verifiability article maintenance templates, distribution of sources by reliability.\\                            
\midrule

Content quality          &
Ratio of stub articles; editing depth; distribution of articles by community quality grading, ORES scoring~\cite{halfaker2020ores}, number of editors, number of quality flaw templates, distribution of edits by source type (i.e., editor, newly-registered editor, admin, bot, IP).\\
\midrule

Content controversiality &
Ratio of locked articles; distribution of articles by controversiality~\cite{yasseri2014most}, distribution articles by number of comments in discussion page and n-chains in discussion pages~\cite{laniado2011wikipedians}.\\                                                  \midrule
\midrule

Media &
Distribution of mentions/references and visits by online media outlets, social media platforms and search engines.\\
\midrule
\midrule

Geopolitics &
Democratic quality scores derived from views and edits by country and well-established country democratic indexes (e.g., \cite{coppedge2017v,schenkkan2019freedom}). \\
\bottomrule
\end{tabular}
\end{table*}

\begin{figure}[h!]
  \centering
  \includegraphics[trim={0cm -0.15cm 0cm 0}, clip, width=\linewidth]{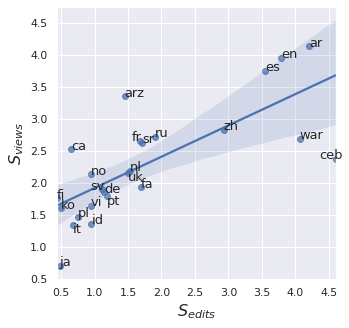}
  \caption{Entropy values ($S$) of the distributions of the number of edits and views by country of the Wikipedia language editions, identified by the ISO 639-1 code, with over 500K articles. The graph includes a linear regression model fit.}
  \label{fig:entropies}
\end{figure}

To illustrate the value of the indicators for knowledge integrity risk assessment in Wikipedia, we provide an example on community demographics, in particular, geographical diversity. Figure~\ref{fig:entropies} shows the entropy value of the distributions of number of edits and views by country of the language editions with over 500K articles. The data has been collected from November 2018 to April 2021. On the one hand, we observe large entropy values for both edits and views in the Arabic, English and Spanish editions, i.e., global communities. On the other hand, other large language editions like the Italian, Indonesian, Polish, Korean or Vietnamese Wikipedia lack that geographical diversity. We should highlight the extraordinarily low entropy of views of the Japanese Wikipedia, which supports one of the main causes attributed to misinformation incidents in this edition~\cite{sato2021}. We also notice the misalignment between high edit entropy and low view entropy values in Cebuano and Waray-Waray editions, which might be the result of the large fraction of content produced by bots distributed around the world~\cite{song2020}. It is also remarkable the misalignment of the Egyptian Arabic Wikipedia with much larger entropy values for views than edits.

\section{Discussion and future work}

This article has presented a preliminary approach to knowledge integrity risk assessment in Wikipedia projects. We have covered the first steps of an ongoing process with the ultimate goal of building a Wikipedia Knowledge Integrity Risk Observatory. The taxonomy relies on risks detected in previous work on knowledge integrity in Wikipedia, nevertheless, it could be enriched through the review of additional literature on risk assessment in other web platforms. 

The analysis shown in this paper has focused exclusively on community demographics. We will extend this work by implementing the rest of the indicators of this and other categories from the taxonomy to assess their informative value. As mentioned above, the current indicators are essentially item counts and distributions of items over features. Future work will also focus on defining advanced metrics while preserving the criteria of ease of interpretation, comparability across wikis and language-agnosticism. Also, indicators should be periodically updatable to allow longitudinal observations. Another future challenge will be to define indicators with finer levels of granularity, that is to say, metrics computed not only on an entire Wikipedia project but on categories, pages, etc.

Last but not least, following the principles of openness, transparency and accountability of Wikimedia~\footnote{\url{https://meta.wikimedia.org/wiki/Wikimedia_Foundation_Guiding_Principles}}, we expect to release the Wikipedia Knowledge Integrity Risk Observatory as a dashboard available to the global movement of volunteers. Therefore, we will also focus on designing an open technological infrastructure to provide Wikimedia communities with valuable information on knowledge integrity.

\bibliographystyle{ACM-Reference-Format}
\bibliography{references}
\end{document}